\documentclass[twoside,twocolumn,9pt]{article}
\usepackage{extsizes}
\usepackage[super,sort&compress,comma]{natbib} 
\usepackage[version=3]{mhchem}
\usepackage[left=1.5cm, right=1.5cm, top=1.785cm, bottom=2.0cm]{geometry}
\usepackage{balance}
\usepackage{mathptmx}
\usepackage{sectsty}
\usepackage{graphicx} 
\usepackage{lastpage}
\usepackage[format=plain,justification=justified,singlelinecheck=false,font={stretch=1.125,small,sf},labelfont=bf,labelsep=space]{caption}
\usepackage{float}
\usepackage{fancyhdr}
\usepackage{fnpos}
\usepackage[english]{babel}
\addto{\captionsenglish}{%
  
}
\usepackage{array}
\usepackage{droidsans}
\usepackage{charter}
\usepackage[T1]{fontenc}
\usepackage[usenames,dvipsnames]{xcolor}
\usepackage{setspace}
\usepackage[compact]{titlesec}
\usepackage{hyperref}

\usepackage{epstopdf}
\usepackage{siunitx}
\usepackage{xspace}
\definecolor{grayCode}{gray}{0.9}

\definecolor{cream}{RGB}{222,217,201}

\begin{document}

\pagestyle{fancy}
\thispagestyle{plain}
\fancypagestyle{plain}{
\renewcommand{\headrulewidth}{0pt}
}

\makeFNbottom
\makeatletter
\renewcommand\LARGE{\@setfontsize\LARGE{15pt}{17}}
\renewcommand\Large{\@setfontsize\Large{12pt}{14}}
\renewcommand\large{\@setfontsize\large{10pt}{12}}
\renewcommand\footnotesize{\@setfontsize\footnotesize{7pt}{10}}
\makeatother

\renewcommand{\thefootnote}{\fnsymbol{footnote}}
\renewcommand\footnoterule{\vspace*{1pt}%
\color{cream}\hrule width 3.5in height 0.4pt \color{black}\vspace*{5pt}} 
\setcounter{secnumdepth}{5}

\makeatletter 
\renewcommand\@biblabel[1]{#1}            
\renewcommand\@makefntext[1]%
{\noindent\makebox[0pt][r]{\@thefnmark\,}#1}
\makeatother 
\renewcommand{\figurename}{\small{Fig.}~}
\sectionfont{\sffamily\Large}
\subsectionfont{\normalsize}
\subsubsectionfont{\bf}
\setstretch{1.125} 
\setlength{\skip\footins}{0.8cm}
\setlength{\footnotesep}{0.25cm}
\setlength{\jot}{10pt}
\titlespacing*{\section}{0pt}{4pt}{4pt}
\titlespacing*{\subsection}{0pt}{15pt}{1pt}

\fancyfoot{}
\fancyfoot[LO,RE]{\vspace{-7.1pt}\includegraphics[height=9pt]{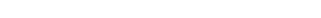}}
\fancyfoot[CO]{\vspace{-7.1pt}\hspace{13.2cm}\includegraphics{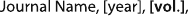}}
\fancyfoot[CE]{\vspace{-7.2pt}\hspace{-14.2cm}\includegraphics{head_foot/RF}}
\fancyfoot[RO]{\footnotesize{\sffamily{1--\pageref{LastPage} ~\textbar  \hspace{2pt}\thepage}}}
\fancyfoot[LE]{\footnotesize{\sffamily{\thepage~\textbar\hspace{3.45cm} 1--\pageref{LastPage}}}}
\fancyhead{}
\renewcommand{\headrulewidth}{0pt} 
\renewcommand{\footrulewidth}{0pt}
\setlength{\arrayrulewidth}{1pt}
\setlength{\columnsep}{6.5mm}
\setlength\bibsep{1pt}

\makeatletter 
\newlength{\figrulesep} 
\setlength{\figrulesep}{0.5\textfloatsep} 

\newcommand{\topfigrule}{\vspace*{-1pt}%
\noindent{\color{cream}\rule[-\figrulesep]{\columnwidth}{1.5pt}} }

\newcommand{\botfigrule}{\vspace*{-2pt}%
\noindent{\color{cream}\rule[\figrulesep]{\columnwidth}{1.5pt}} }

\newcommand{\dblfigrule}{\vspace*{-1pt}%
\noindent{\color{cream}\rule[-\figrulesep]{\textwidth}{1.5pt}} }

\makeatother

\twocolumn[
  \begin{@twocolumnfalse}
{\includegraphics[height=30pt]{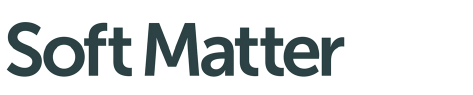}\hfill\raisebox{0pt}[0pt][0pt]{\includegraphics[height=55pt]{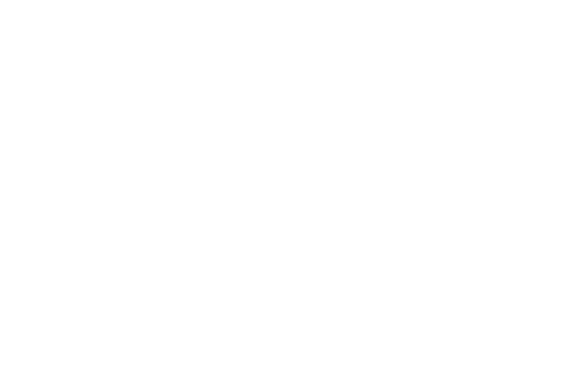}}\\[1ex]
\includegraphics[width=18.5cm]{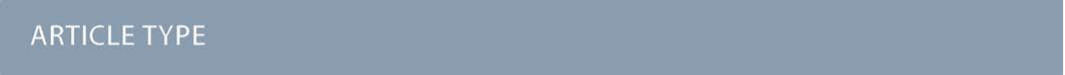}}\par
\vspace{1em}
\sffamily
\begin{tabular}{m{4.5cm} p{13.5cm} }
\includegraphics{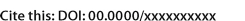} & \noindent\LARGE{\textbf{Pattern Formation in Crumpled Hydrogel upon Rapid Dehydration with Acetone$^\dag$}} \\
\vspace{0.3cm} & \vspace{0.3cm} \\

 & \noindent\large{G. T. Fortune$^{a}$, M. A. Etzold$^{b}$, J. R. Landel$^{c,d}$ and Stuart B. Dalziel$^{a}$} \\

\includegraphics{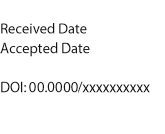} & \noindent\normalsize{

From microactuators to biological tissues, non-porous materials with the ability to strongly expand when in contact with a solvent are ubiquitous. Consequently, the swelling of polymer systems such as hydrogel has received recently much research attention. However, the related dehydration of these systems has received much less attention. Here, we present experiments investigating the rapid dehydration of a swollen hydrogel sheet whose surface exhibits a transient crumpling instability characterised by line segments of cusps patterning the surface of the gel into an array of bumps. We perform this dehydration through immersion in acetone, which is highly miscible in water, but poorly miscible in the hydrogel. We report the onset of a fascinating pattern formation where regions of the hydrogel sheet turn turbid. We find that the emerging pattern is independent of the overall extent of the hydrated swollen surface. The pattern wavelength only depends on the duration of hydration before immersion in acetone, growing temporally with a power law behaviour. We conclude through drawing comparisons between features of this dehydration induced pattern and the original crumpling instability in the water swollen hydrogel sheet.
} \\

\end{tabular}

 \end{@twocolumnfalse} \vspace{0.6cm}

  ]

\renewcommand*\rmdefault{bch}\normalfont\upshape
\rmfamily
\section*{}
\vspace{-1cm}


\footnotetext{\textit{$^{a}$~Department of Applied Mathematics and Theoretical Physics, Centre for Mathematical Sciences, University of Cambridge, Wilberforce Road, Cambridge CB3 0WA, United Kingdom.}}
\footnotetext{\textit{$^{b}$~The Defence Science and Technology Laboratory, Porton Down, B383B Room 1/6, Salisbury, Wiltshire, SP4 0JQ. }}
\footnotetext{\textit{$^{c}$~Universite Claude Bernard Lyon 1, Laboratoire de Mécanique des Fluides et d'Acoustique (LMFA), UMR5509, CNRS, Ecole Centrale de Lyon, INSA Lyon, 69622 Villeurbanne, France. }}
\footnotetext{\textit{$^{d}$~Department of Mathematics, Alan Turing Building, University of Manchester, Oxford Road, Manchester, M13 9PL, UK. }}
\footnotetext{\dag~Supplementary movies are available at DOI: 10.1039/cXsm00000x/}


\section{Introduction}
Hydrogels are a category of polymeric gels assembled from cross-linked polymer molecules in water that form three-dimensional networks \cite{Laftah11} of macroscopic extent \cite{Vervoort06}. Large volumetric strains, induced by swelling in response to changing environmental conditions, can lead to an interdisciplinary class of problems, where chemically-driven transport within the polymer drives large changes in the geometry of the hydrogel \cite{Guvendiren10,Etzold21}. Macrosocopically these systems exhibit poroelastic behaviour. On short time scales \cite{Yoon10} (e.g. seconds) hydrogels behave like hyperelastic incompressible solids. On longer timescales (e.g. minutes or hours), hydrogels display more fluid like behaviour, responding to changing external conditions (stress, temperature, humidity) by swelling, shrinking or deforming.

Some hydrogels can absorb very large quantities of water, up to thousands of times their dry weight in water. This water absorption capacity, together with the fact that hydrogels have similar physico-chemical characteristics as many tissues \cite{Drury03,Hoffman12}, make hydrogels an excellent substitute for tissues in the laboratory (e.g. when studying decompression sickness \cite{Walsh17,Zhang21}). For over sixty years since the pioneering work of \citet{Wichterle60}, who developed the first contact lens using a novel pHEMA bio-compatible hydrogel, hydrogels have been extensively used in a broad range of  medical contexts \cite{Hoffman12}. For instance, they are used in tissue engineering  as a matrix framework for the repair and regeneration of a wide range of tissues and organs (\citet{Lee01} gives a general review while \citet{Tang20} and \citet{Madhusudanan20} present applications in the spinal column and brain, respectively), or in wound dressing where they maintain a moist healing environment allowing gaseous exchange, and thus promote wound healing \cite{Deng22,Stubbe21,Tortorella21}.

Since the mechanical properties of hydrogels can be widely and easily adjusted through changing the polymer chemistry or cross-linker density \cite{Ozcelik16,Tortorella21}, high-swelling hydrogels have also found wide-spread applications beyond medicine. The ability of external stimuli to induce swelling that yields large deformation has been harnessed to drive microfluidic pumps \cite{Kwon11,Richter09}, and has led to their use as micro-actuators for optical control, flow control, sensor and micro-robotics applications \cite{Ionov14,Porter07}. Understanding the connection between swelling mechanics and the microscopic driving forces is crucial for the design and optimisation of these systems.

\begin{figure}[ht]
\centering
  \includegraphics[height=4cm]{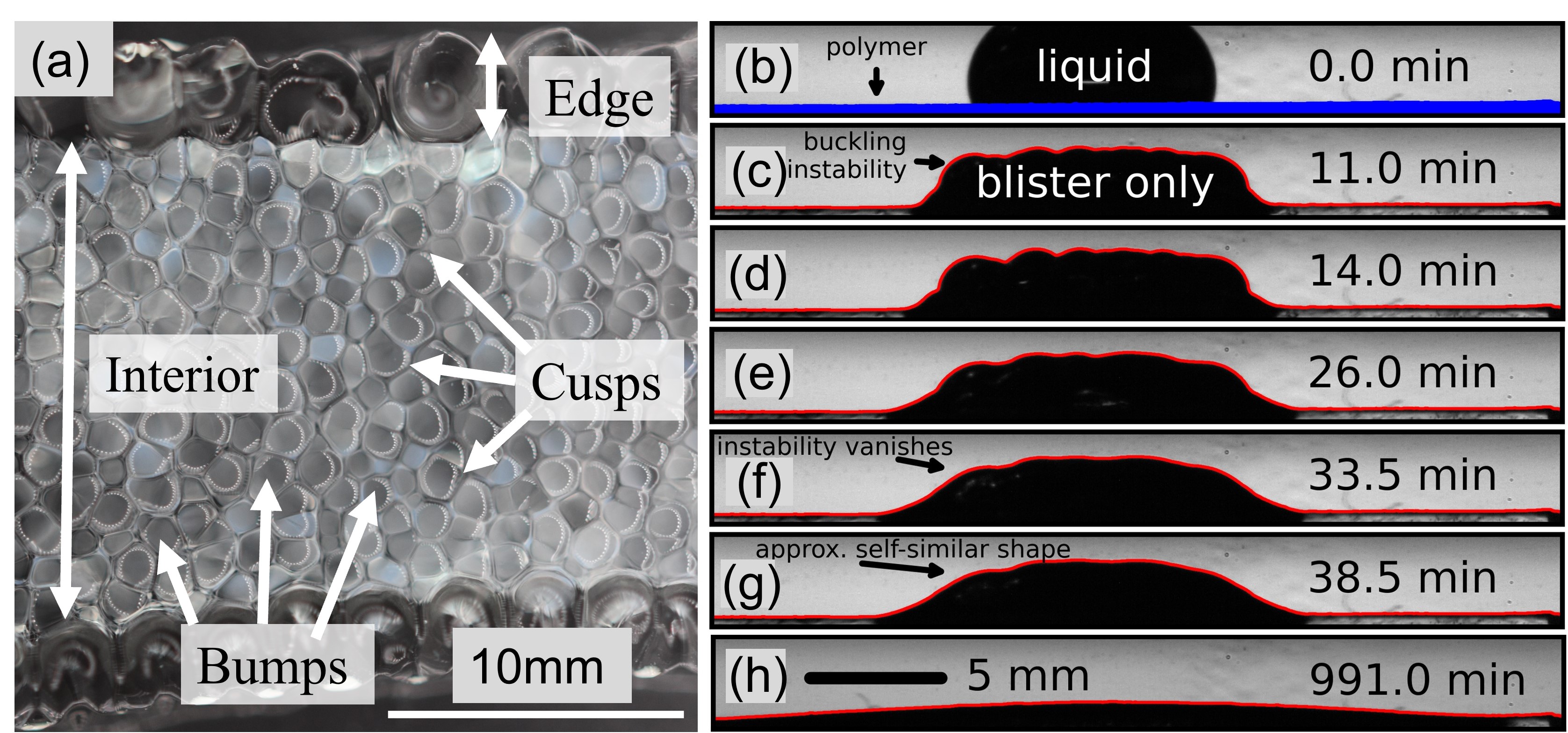}
  \caption{Crumpling instability that emerges when hydrogel swells with water. (a) Instability across the entirety of a \SI{48}{\milli\metre} by \SI{16}{\milli\metre} hydrogel sheet of thickness $\approx \SI{1}{\milli\metre}$ that has been immersed in water for five minutes. Note that the edges swell considerably more than the interior. (b--h) Instability on a blister formed by the absorption of a \SI{100}{\mu\litre} water drop by a hydrogel sheet of horizontal dimensions \SI{16}{\milli\metre} by \SI{16}{\milli\metre} immersed in oil. Adapted from figure 2 of \citet{Etzold23}. (b) Initial configuration at deposition of the water droplet. (c--d): The surface instability forms. (e--g): Transition towards the long-time spreading regime which is shown in (h). The visible hydrogel sheet is marked in blue in (b). From (c) onward a red line marks the hydrogel--oil interface.}
  \label{fig:Fortune1}
\end{figure}

It is known that when hydrogels undergo extensive rapid swelling, a transient crumpling instability forms on the surface of the gel \cite{Tanaka87} characterised by many line segments of cusps patterning the gel surface into an array of bumps. 
Figure \ref{fig:Fortune1} shows two examples of this instability, forming at the surface of a \SI{1}{\milli\metre} thick hydrogel sheet: in (a) the sheet has been completely immersed in water, in (b--h) a \SI{100}{\micro\litre} water drop has been deposited at the surface of the sheet, which, upon absorption, forms a blister where the instability appears at early time (c--f). As can be seen, the pattern consists of many line segments of cusps into the gel that arise from shear bending of a homogeneously swollen gel surface. 

Initially, only a very thin superficial layer of the gel substrate is swollen at the water--gel interface. The inner part of this layer is fixed to the core of the unswollen gel while the outer part is free to expand. This generates stress gradients in the swollen layer, leading to an osmotic pressure difference across the interface. When this pressure difference is small, the hydrogel stretches in the direction perpendicular to the surface. When this pressure difference becomes larger, the outer surface is forced to buckle \cite{Tanaka87}. 

This pattern starts with a small wavelength, proportional to the thickness of the swollen layer. As time increases, the thickness of the swollen layer increases, leading to an increase in the wavelength of the pattern. Finally, when the wavelength becomes comparable to the horizontal gel lengthscales, i.e. the gel width and length, the instability gradually disappears. It is believed that all gels, provided that they swell strongly enough, both arising from man-made or natural polymers, can form this instability \cite{Tanaka87}. 

In this article, we demonstrate that when a hydrogel sheet, whose upper surface exhibits a crumpling instability, is rapidly dehydrated (e.g. through immersion in acetone), a pattern formation occurs with regions of the hydrogel changing colour from transparent to turbid. We explore and report experimental observations about this instability, analysing both the pattern before immersion in acetone and after. 

In $\S$\ref{experiments}, we describe experiments tracking the absorption of water by a thin layer of hydrogel and subsequent dehydration with an excess of acetone. The results of these experiments are described in \S\ref{observations}, in particular the patterns formed upon dehydration.  
The remainder of the paper is spent investigating this behaviour. In particular in \S\ref{ConstantdehydrationTime}, we demonstrate that the pattern that emerges is independent of the total size of the blister, rather just a function of the time before dehydration with acetone. Furthermore, in \S\ref{ConstantVolume}, we show that the temporal evolution of the wavelength of the pattern behaves as a power law. A comparison with additional experiments exploring the original crumpling instability in water swollen hydrogel indicates that the pattern wavelength is set by the wavelength of the crumpling instability at the point in time when acetone is added to the system. We conclude the paper in \S\ref{conclusions} by discussing the underlying physics that control this instability mechanism. 

\section{Experiments} \label{experiments}
The first phase of the experiments studied here correspond to the first stages of the experiments performed by  Etzold et al. \cite{Etzold23}, who focused their study on the long-time evolution of the blister, when the instability has disappeared and the blister is self-similar. In what follows below, we refer to the time once the water is added before the subsequent swelling is quenched by the addition of excess acetone as the hydration duration.

\subsection{Materials and Methods} \label{methods}
We utilised commercially available medical-grade hydrogel pads (Hydrogel Nipple Pads, Medela, Switzerland), which were manufactured as approximately $8\times\SI{8}{\centi \metre \squared}$ sheets. Due to their proprietary nature, the full details of their synthesis are not known. The hydrogel pads originally appear as a rubbery sheet with a slightly tacky surface, namely the material has a high polymer content in a state known as a `rubbery state' \cite{Etzold23}. After prolonged exposure to the atmosphere (ca. 25-\SI{40}{\percent} relative humidity, ca. \SI{22}{\celsius}), their initial thickness was determined with a micrometer as $a_0=\SI{1.13}{\milli \metre}\pm\SI{0.01}{\milli \metre}$. One side of a sheet was fixed to a thin plastic sheet (impermeable to water) of approximate thickness \SI{0.04}{\milli \metre}. The other side was protected by a thicker and much more easily removed plastic film. The hydrogel sheets incorporated a thin non-woven gauze during their manufacture. This was hypothesised to provide strength in the plane of the sheet. However, this gauze did not appear to affect the swelling of the sheets within the degree of swelling relevant for this work. 

Utilising Karl--Fischer titration, a standard volumetric titration method to determine trace
amounts of water in a sample, the water content of a typical hydrogel pad that had equilibrated
with the laboratory air was measured to be $22\%$ \citep{Etzold23}. 
These  sheets were  cut into coupons approximately $16\times\SI{16}{\milli \metre \squared}$. Utilising pliers, both the plastic film and the plastic sheet were removed from each coupon, taking care to not disturb or stretch the hydrogel. The coupons were then placed in the middle of \SI{60}{\milli\metre} diameter soda-lime glass Petri dishes with the side of the sheet that had been in contact with the thin plastic sheet in contact with the surface of the Petri dish. Residual adhesive left from how the plastic sheet had been bonded onto the hydrogel sheet was found to be sufficient to stop lateral movement of the sheet. Each Petri dish containing a sheet was then left for at least an hour open to the atmosphere to allow for atmospheric equilibration. 

Deionised water was then added to the hydrogel sheets dropwise utilising micro pipettes (Thermo Scientific™ Finnpipette™ F2 Variable 2--20 \si{\mu\litre} and 20--200 \si{\mu\litre} Volume Pipettes). The drops deposited at the surface of the hydrogel were mostly sessile drops with pinned contact line and roughly circular shape, with typical horizontal extent roughly 1 \si{\cm} and vertical thickness a few millimetres. The Petri dishes were then sealed with a Petri dish lid before being left for given hydration periods. At the end of the hydration periods, the Petri dishes were filled with an excess of lab grade acetone ($\approx \SI{10}{\milli\litre}$). After dehydration through immersion in acetone, the coupons were left for at least fifteen minutes to observe any potential subsequent evolution of the pattern. We observed no changes in the pattern at the surface of the hydrogel from the moment of immersion in acetone. This showed that the acetone acts very rapidly to  stabilize and lock the pattern. Hence, all subsequent measurements and imaging were performed without removing the acetone and regardless of the time duration after immersion in acetone.
\begin{figure}[!hb]
	\centering\includegraphics[trim={0 0cm 0 0cm}, clip, width=\columnwidth]{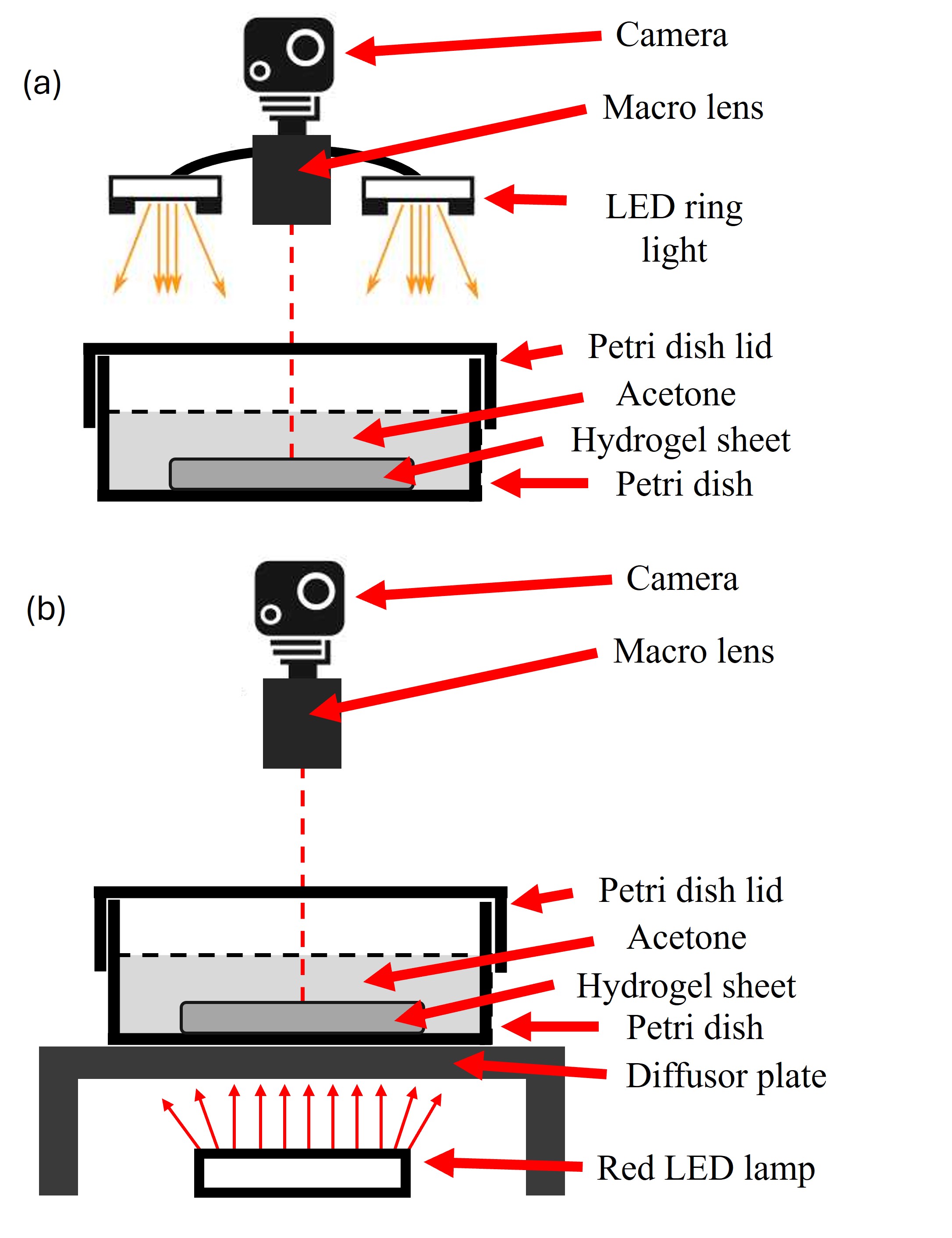}
	\caption{Schematics of the laboratory apparatus to study the absorption of water into a thin hydrogel sheet and its subsequent dehydration in acetone, using: (a) white light illumination from above; and, (b) red light illumination from below.}
	\label{fig:Fortune2}
\end{figure}

\subsection{Imaging system and data analysis}
As illustrated by schematics in Fig. \ref{fig:Fortune2}, a Nikon D3500 camera with a AF-S Micro NIKKOR 60mm f/2.8G macro lens, mounted vertically downwards using a Manfrotto tripod with reversible centre post, recorded the evolution of the pattern at the surface of the hydrogel sheets. We used either lighting from above using a lens-mounted white LED ring light (Fig. \ref{fig:Fortune2}(a), Aputure Amaran Halo), or from below using a red LED light source (Fig. \ref{fig:Fortune2}(b), custom-made circuit board containing a four by six array of red SMD LEDs with centre-wavelength \SI{630}{\nano \metre}, 0.5A, 12V).

Image analysis was utilised to extract the locations of the edges of the cells of the pattern in order to generate and analyse quantitative statistical data about the instability pattern. This combined both manual steps using the open-source image processing package Fiji \cite{Schneider12,Schindelin12} and automated steps using several custom-made MATLAB scripts that utilised MATLAB's Image Processing Toolbox \cite{Matlab20}.

\subsection{Additional qualitative experiments} \label{waterdunking}
For qualitative comparison, an additional set of experiments was performed to visualise the crumpling instability on hydrogel sheets that had only been swelled in water (i.e. no immersion in acetone). Large hydrogel coupons, approximately $\SI{16}{\milli\metre}\times\SI{48}{\milli \metre}$, were prepared using the  procedure  described in \S\ref{methods}, and placed in the middle of Petri dishes. The Petri dish was filled with an excess of deionised water and then the hydrogel was left to swell for thirty seconds. The sheet was removed, placed onto the surface of a second, empty, Petri dish, and then imaged using the setup showed in Fig. \ref{fig:Fortune2}. The sheet was then repeatedly placed in the water bath to swell for thirty seconds before removal and imaging. Care was taken to minimise loss of hydrogel during the transfer steps. This method highlights the changing topographic features of the hydrogel--air interface very well but is unable to yield quantitative measurements of the temporal evolution of the wavelength since the hydrogel sheet is not continuously swollen by the water bath. The hydrogel sheet was not imaged continuously when immersed since the spatial variations in refractive index hindered visualisation of topographic detail.

\subsection{Additional quantitative experiments} \label{waterdunking2}
A second additional set of experiments was performed to obtain quantitative measurements of the temporal evolution of the wavelength of the crumpling instability in hydrogel that had just been swelled in water. As in \S\ref{waterdunking}, larger hydrogel coupons were prepared and placed in Petri dishes. These Petri dishes were then filled with an excess of deionised water that contained a high concentration of a blue vegetable food dye (Ingram Brothers Dark Blue Food Colouring, E122 + E133). The vegetable dye, being made of large molecules, was observed to not absorb into the hydrogel. Hence, the dye remained at the top of the hydrogel, predominantly due to gravity in the creases of the crumpling instability and at most we believe penetrating just a few molecules thick into the hydrogel. Imaging the evolution of this system from above with a similar setup to that shown in Fig. \ref{fig:Fortune2}(b), namely using a red LED lamp as the light source, the bumps of the instability appeared white while the creases appeared black. Hence, with suitable post processing in MATLAB (thresholding the images to just retain the bumps), the average wavelength of the instability could be quantified. 

\subsection{Experimental observations} \label{observations}
\begin{figure}
	\centering\includegraphics[trim={0 0cm 0 0cm}, clip, width=\columnwidth]{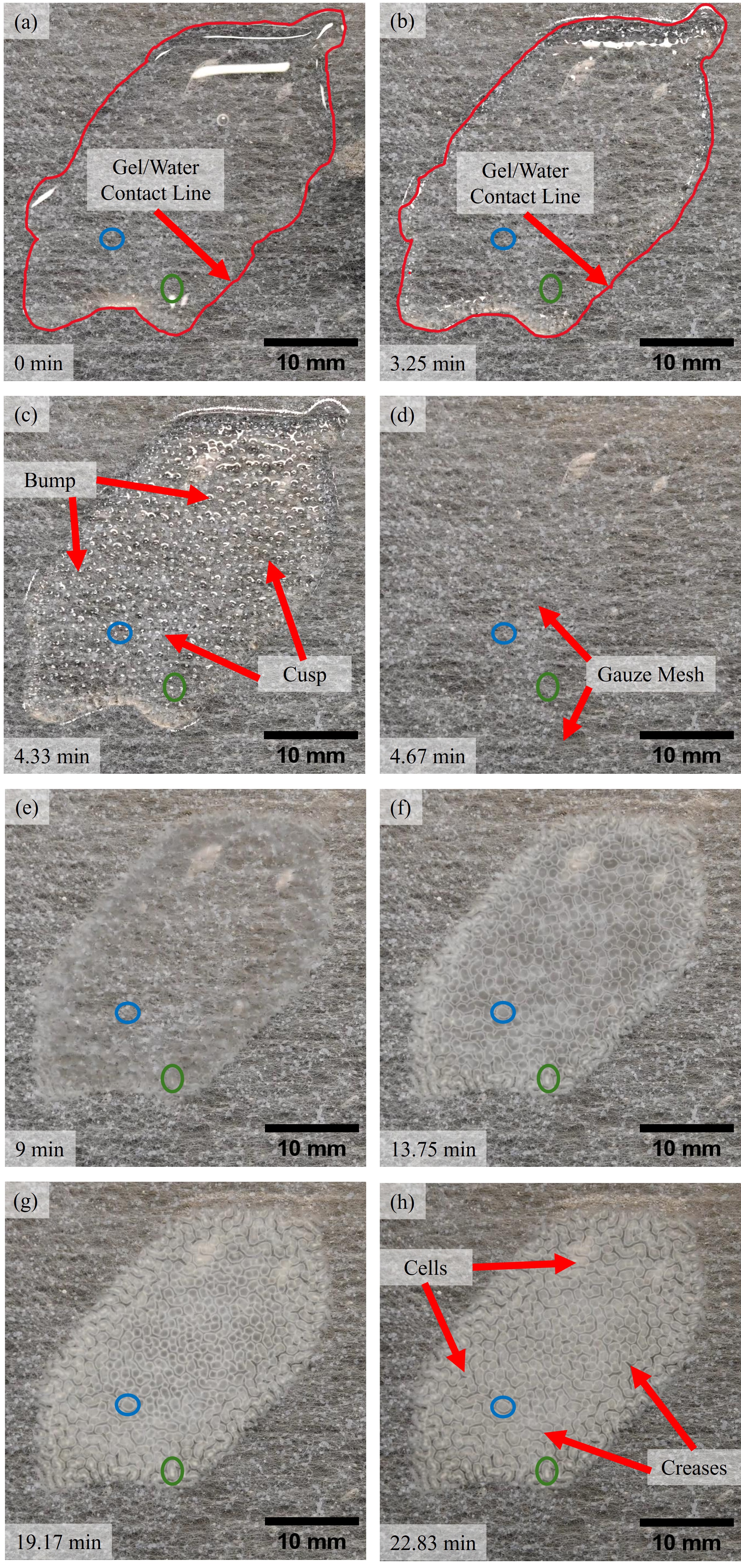}
	\caption{Sequence of top-view photographs showing the absorption of water into a hydrogel sheet followed by subsequent dehydration through immersion in acetone. (a-b) States with liquid water remaining. The transient apparent edge of the water droplet is drawn in red for ease of visualization.  (c) The surface crumpling instability forms. (d) Acetone is added. (e-h) 'Curing' stage as the pattern of white (turbid)  and transparent shapes appears at the surface of the hydrogel. Blue and green circles in (a-h) track the evolution of two regions which initially contain bumps of the crumpling instability before transitioning into cells of the acetone induced pattern, one at the edge of the blister (green) and one in the interior of the blister (blue). 
	}
	\label{fig:Fortune3}
\end{figure}
A montage of typical experimental images is shown in Fig. \ref{fig:Fortune3}. Upon contact with the hydrogel surface, the water droplet sat on top of the hydrogel surface with a large contact angle at the water--air--hydrogel contact line (Fig. \ref{fig:Fortune3}a). We note that the concept of a triple phase contact line is not well-defined here, since the system is not at equilibrium. Hence, our use of the concise phrase `contact line' is in principle a slight abuse of language to describe the transient location of where the edge of the water droplet appears at the surface of the hydrogel at early times in the experiments. When the water reached a sessile drop state on the hydrogel surface with the contact line advancing slightly, the surface water and swollen hydrogel coexisted as the water absorbs into the hydrogel to form a  blister (Fig. \ref{fig:Fortune3}(b)). Note that for ease of visualization, the contact line has been drawn in red in Figs. \ref{fig:Fortune3}(a,b).
Once all the surface water had been absorbed, a surface instability appeared, forming a blister with a crumpled surface characterised by a pattern of bumps and cusps (Fig. \ref{fig:Fortune3}(c)).

In the next stage, the hydrogel is rapidly dehydrated through the addition of an excess of acetone. Initially, the crumpling pattern disappears as water desorbs out of the hydrogel due to a difference of chemical potentials at the interface (Fig. \ref{fig:Fortune3}(d)). This chemical potential difference arises from the polarity of the carbonyl group of acetone leading to the complete miscibility of water and acetone. However, over the course of the next 15 minutes, a pattern of optically opaque cells which appear turbid under visible light form on the surface of the hydrogel (Figs. \ref{fig:Fortune3}(e--h)). It is important to note that the hydrogel sheet surface is at this point completely flat i.e. no surface topography. Between these cells are optically transparent regions that we  denote as creases. The location of the cells and creases qualitatively match the location of the bumps and cusps observed in the original crumpling instability, before immersion in acetone. Away from the blister, the hydrogel remains optically transparent, noting that the thin gauze mesh described in above in \S \ref{methods} is visible. For reference, the mesh is pointed out in Fig. \ref{fig:Fortune3}(d). When the hydrogel sheet is cut open with a scalpel, the turbid region is observed to extend a distance into the depth of the sheet on the order of magnitude of half of the hydrogel sheets thickness ($\sim \SI{0.5}{\milli\metre}$), extending deeper into the sheet in experiments with longer times before dehydration. When immersed in acetone, these structures are very stable, lasting many days. However, this process is reversible. If the acetone layer is removed and a water drop is placed again on the blister, the original rubbery hydrogel sheet is recovered. Indeed, just removing the acetone layer is enough to recover the original state on a slightly longer timescale by redistribution due to water vapour phase transport from ambient humidity. 

The evolution of two regions which first contain a bump of the crumpling instability before transitioning into a cell of the acetone induced pattern (one enclosed by a blue circle in the interior of the blister and one enclosed by a green circle at the edge of the blister) are tracked across Figs. \ref{fig:Fortune3}(a--h). We can notice that regions at the edge of the blister become cloudy first. Furthermore, for a given cell, the outer parts closest to the cusps turn cloudy first before the inner regions, leading to a gradient in turbidity across a cell.

\section{Results}

\subsection{Experiments at constant hydration duration}\label{ConstantdehydrationTime}

\begin{figure}[!tb]
	\centering\includegraphics[trim={0 0cm 0 0cm}, clip, width=\columnwidth]{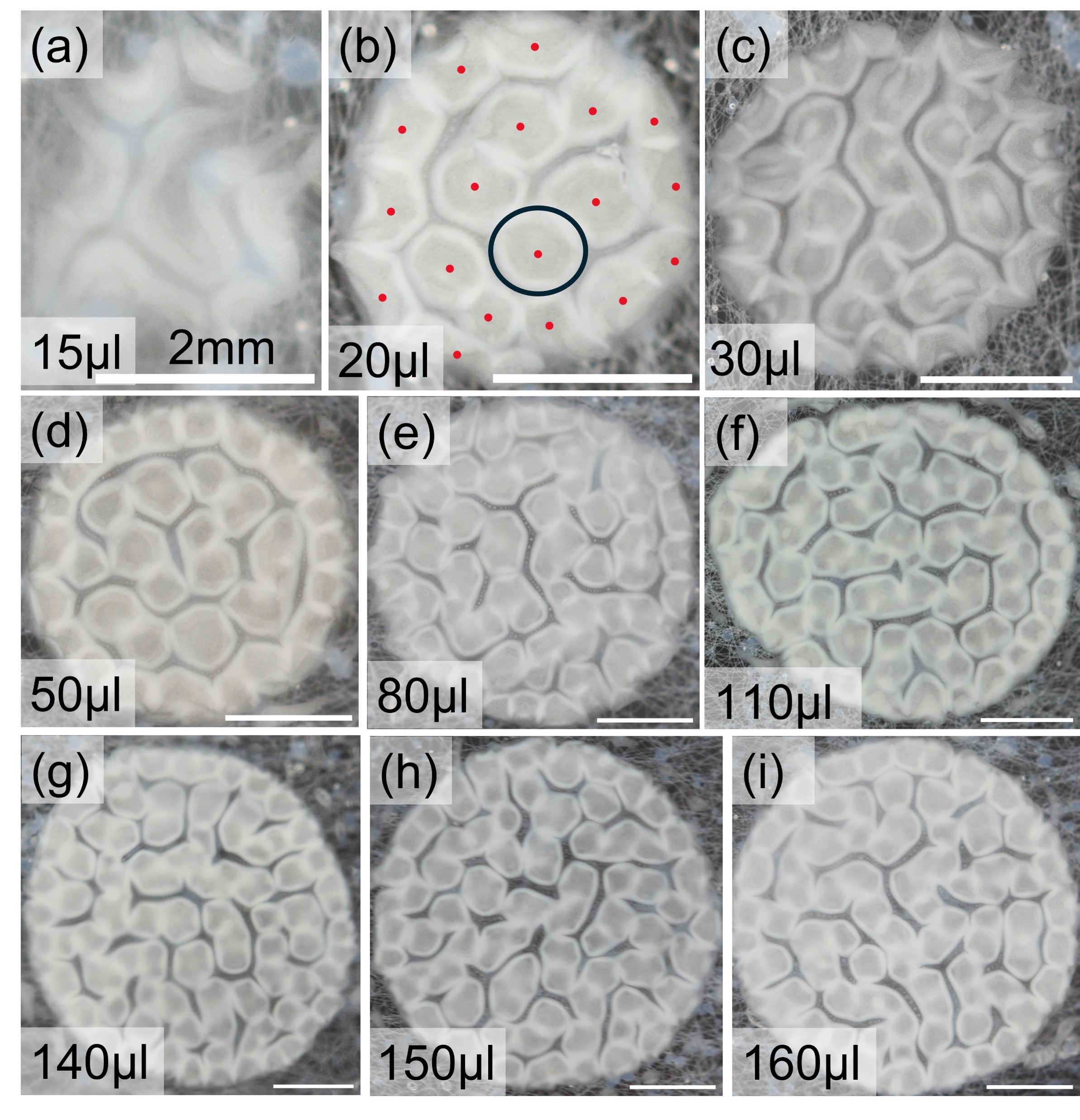}
	\caption{Photographs showing the greyscale pattern of the instability that emerges when a hydrogel sheet that has been swollen through the absorption of a water drop is dehydrated in a static acetone bath after a hydration duration of five minutes for a range of different initial drop volumes. (a) At small  drop volume, the pattern is not clearly defined due to edge effects being pre-dominant across the swollen area. (b--i) Larger drop volumes show the pattern more clearly in the interior of the swollen region. The wavelength and overall structure of the pattern does not appear to depend on the drop volume. The number of cells increases with increasing  drop volume, owing to an increase in the area of the swollen region. All white scale bars denote the same length scale (\SI{2}{\milli\metre}). The red dots in (b) indicates the centers of cells while the black ellipse is fitted to the edges of a cell. 
	}
	\label{fig:Fortune4}
\end{figure}

\begin{figure}[!tb]
	\centering\includegraphics[trim={0 0cm 0 0cm}, clip, width=\columnwidth]{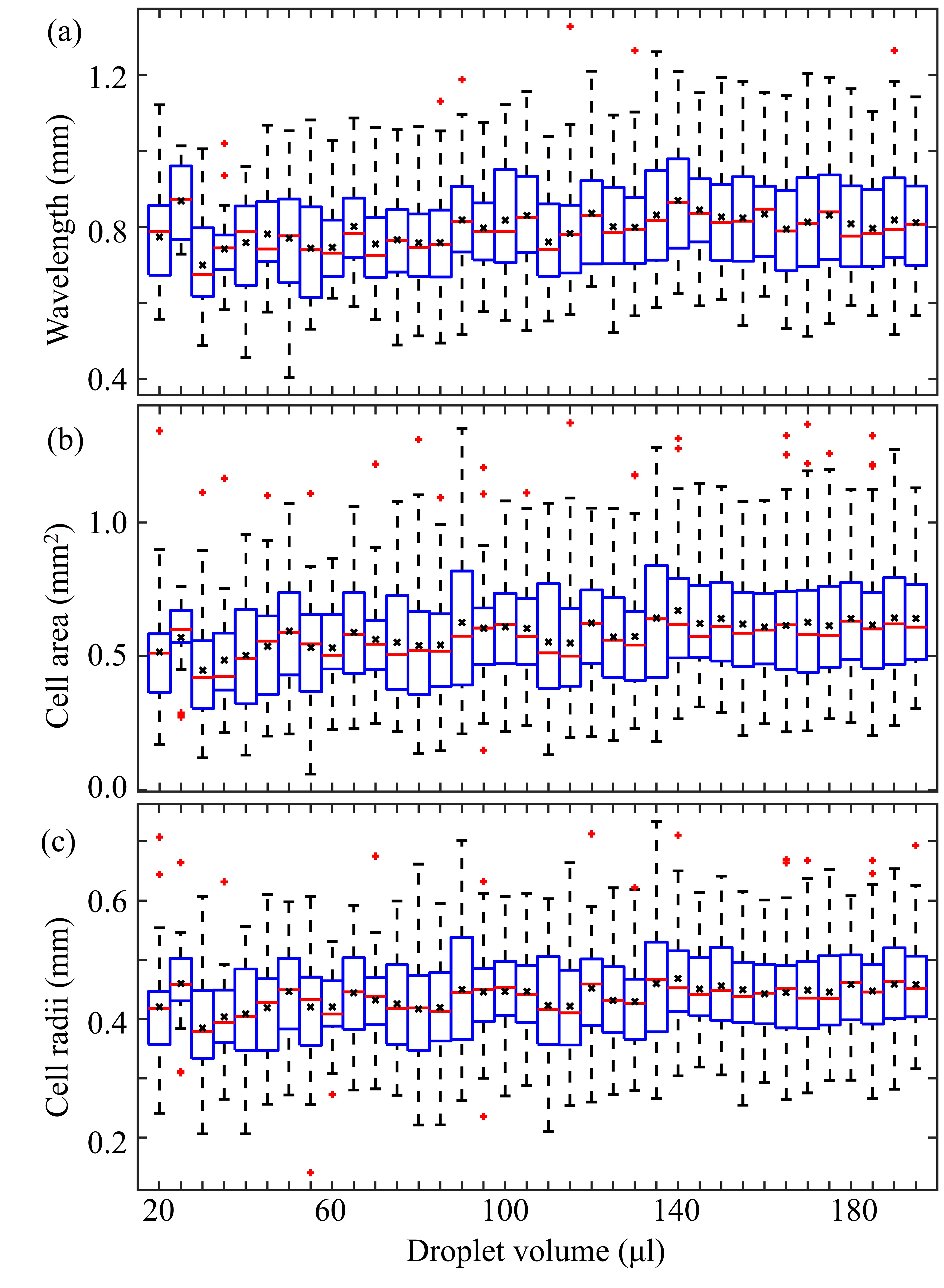}
	\caption{Box plots displaying the variation in (a) pattern wavelength, (b) cell area and (c) cell radius for a range of experiments with a fixed hydration duration of 5 minutes but for different initial water drop volumes. Each entry denotes a different experiment. Each blue box contains the data points for that particular experiment from the 25\textsuperscript{th} percentile to the 75\textsuperscript{th} percentile. The median and mean are denoted by a horizontal red line and a black cross, respectively. The whiskers are defined as including data up to one and a half times the inter-quartile range away from the blue boxes. Any data point that is beyond the whiskers is assumed to be an outlier and denoted by a red plus sign. }
	\label{fig:Fortune5}
\end{figure}

First, we investigate whether the geometrical characteristics of the instability pattern depends on the volume and geometry of the initial water volume placed onto the hydrogel sheet. Fig. \ref{fig:Fortune4} presents photographs showing the patterns that emerge when a hydrogel sheet that has been swollen through the absorption of a water drop is dehydrated in a static acetone bath after a fixed hydration duration of five minutes and for a range of different initial water drop volumes. 
Care was taken when placing the water drops onto the sheets to obtain axisymmetric sessile drops which are pinned to the hydrogel surface. However, as can be seen from the images in Fig. \ref{fig:Fortune4}, some asymmetry did still occur. This hydration duration of five minutes was chosen based on a range of preliminary experiments with different hydration durations. We found that a hydration duration of 5 minutes was short enough to produce clear and well-defined patterns, whilst being long enough to yield clearly visible individual cells (a typical cell is shown with a black ellipse in Fig. \ref{fig:Fortune4}b). 

For small  drop volumes, below $\SI{20}{\mu\litre}$, the emerging patterns are not well defined with the cells becoming longer and thinner (see Fig. \ref{fig:Fortune4}a). For small droplets, the pattern seems to be dominated by edge effects where the buckling instability is much less regular. We performed three experiments with smaller droplets ($\SI{2}{\mu\litre}$, $\SI{5}{\mu\litre}$ and $\SI{10}{\mu\litre}$; not shown here) where no pattern emerged at all and very little turbidity, if at all, was observed. We hypothesise that this is due to the fact that the wavelength of the instability is comparable with (or larger than) the radius of the blister. 

For larger initial drop volumes, the resulting patterns are clear and well defined. From visual inspection of the photographs in \ref{fig:Fortune4}(b--i), the pattern appears to be independent of the surface area covered by the  drop, both in terms of the wavelength of the pattern and the geometry of the cells. In Fig. \ref{fig:Fortune5}, we show the distribution of pattern wavelength, cell area and characteristic cell radius for  experiments with the initial droplet volume varying from 10 to $\SI{220}{\mu\litre}$. We define the pattern wavelength as the ensemble average across the blister of the distance from the centre of a cell to its nearest neighbour (the distance from a red dot in Fig. \ref{fig:Fortune4}(b) to its closest red dot neighbour). Furthermore we define the cell radius of a given cell as the average of the major and minor radii of a fitted ellipse to the cell (the black ellipse in Fig. \ref{fig:Fortune4}(b)). There is no evidence of any trend in any of these three metrics as the initial water drop volume changes as the fluctuations are well within experimental uncertainty,  supporting the hypothesis that the pattern that forms is independent of the geometrical properties of the initial water drop. 
\subsection{ Experiments at constant water volume to study the time evolution}\label{ConstantVolume}

\begin{figure}[!tb]
	\centering\includegraphics[trim={0 0cm 0 0cm}, clip, width=\columnwidth]{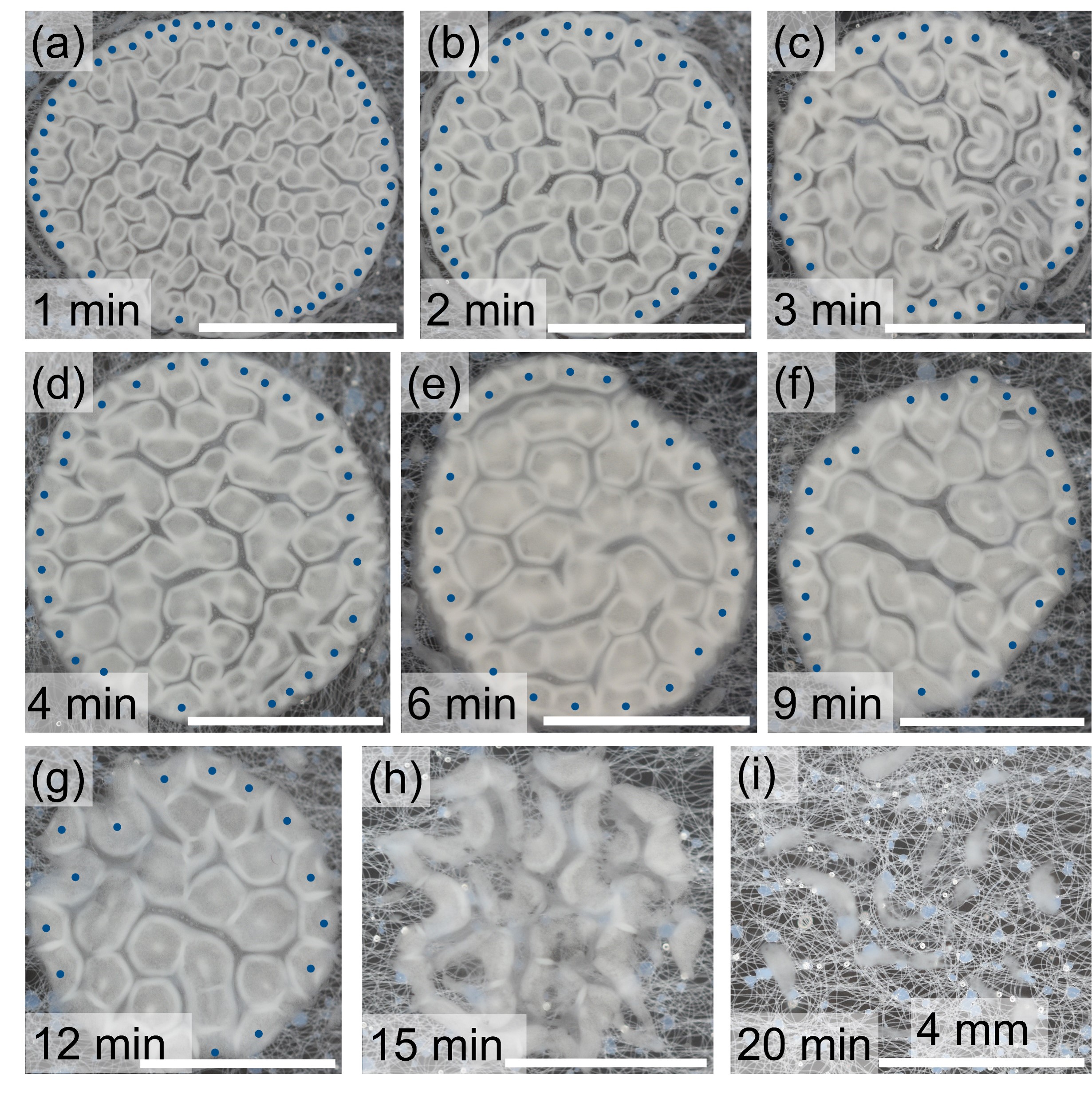}
	\caption{Sequence of photographs showing the patterns that emerge when a hydrogel sheet that has been swollen through the absorbance of a single \SI{100}{\mu\litre} water drop is dehydrated in a static acetone bath for a range of different hydration durations (the time between the addition of the water drop and the addition of the acetone into the system). (a)--(g) The wavelength of the pattern increases with increasing time before dehydration. (h) The pattern becomes less structured and starts to disappear, in particular towards the edge of the blister. (i) The pattern has almost completely disappeared, with only a few small optically opaque regions visible. All white scale bars denote the same distance (\SI{4}{\milli\metre}). In (a--g), the centres of outer cells are denoted by blue dots. 
	}
	\label{fig:Fortune6}
\end{figure}

\begin{figure}[!tb]
	\centering\includegraphics[trim={0 0cm 0 0cm}, clip, width=\columnwidth]{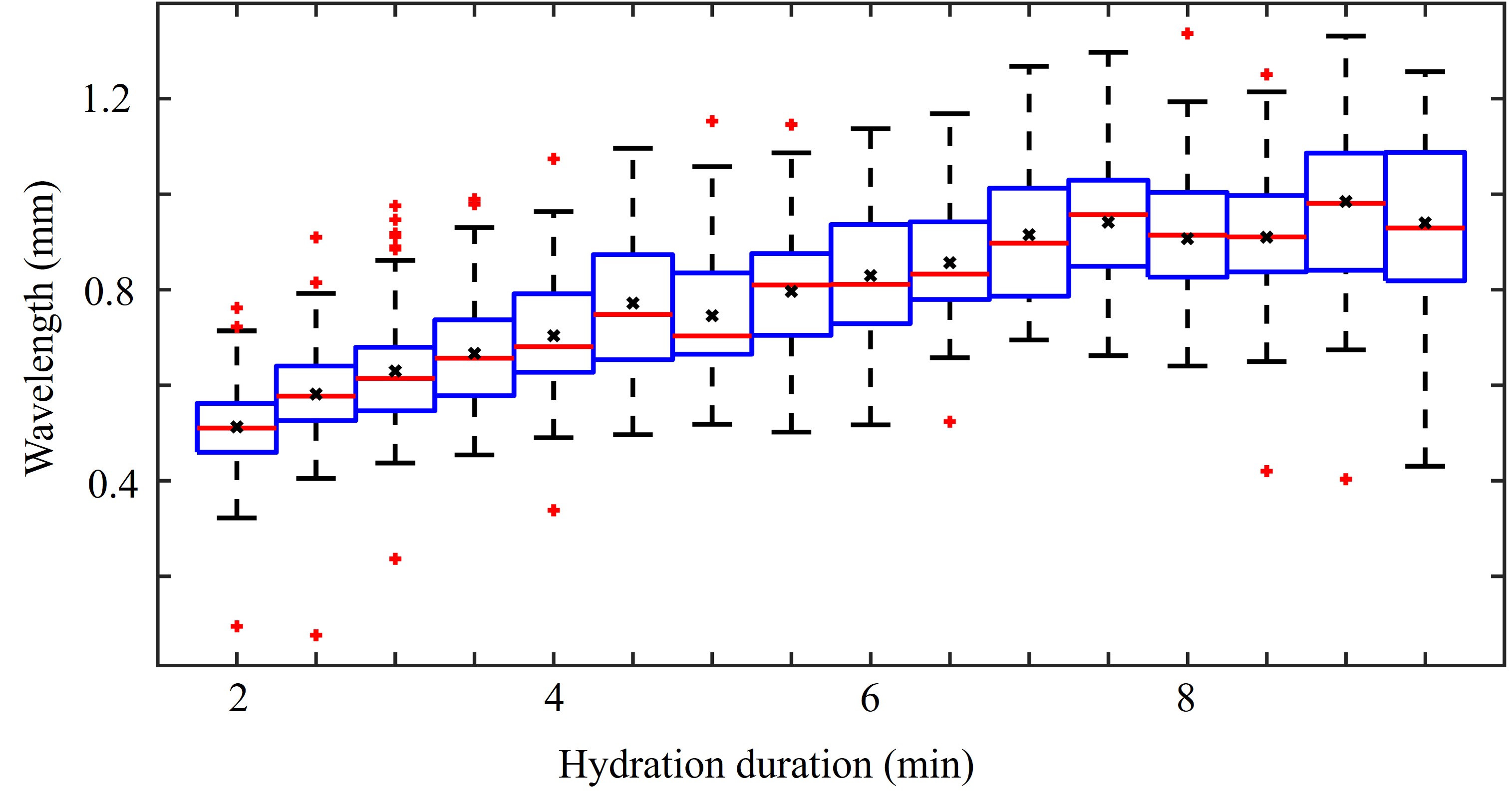}
	\caption{Box plot displaying the variation in pattern wavelength for a set of sixteen experiments with a fixed water drop volume of \SI{100}{\mu\litre} but for increasing hydration durations using hydrogel coupons that came from the same initial hydrogel sheet. Each blue box contains the data points for that particular experiment from the 25\textsuperscript{th} percentile to the 75\textsuperscript{th} centile. The median and mean are denoted by a horizontal red line and a black cross, respectively. The whiskers are defined as including data up to one and a half times the inter-quartile range away from the blue boxes. Any data point that is beyond the whiskers is assumed to be an outlier and denoted by a red plus sign. 
	}
	\label{fig:Fortune7}
\end{figure}

\begin{figure}[!tb]
	\centering\includegraphics[trim={0 0cm 0 0cm}, clip, width=\columnwidth]{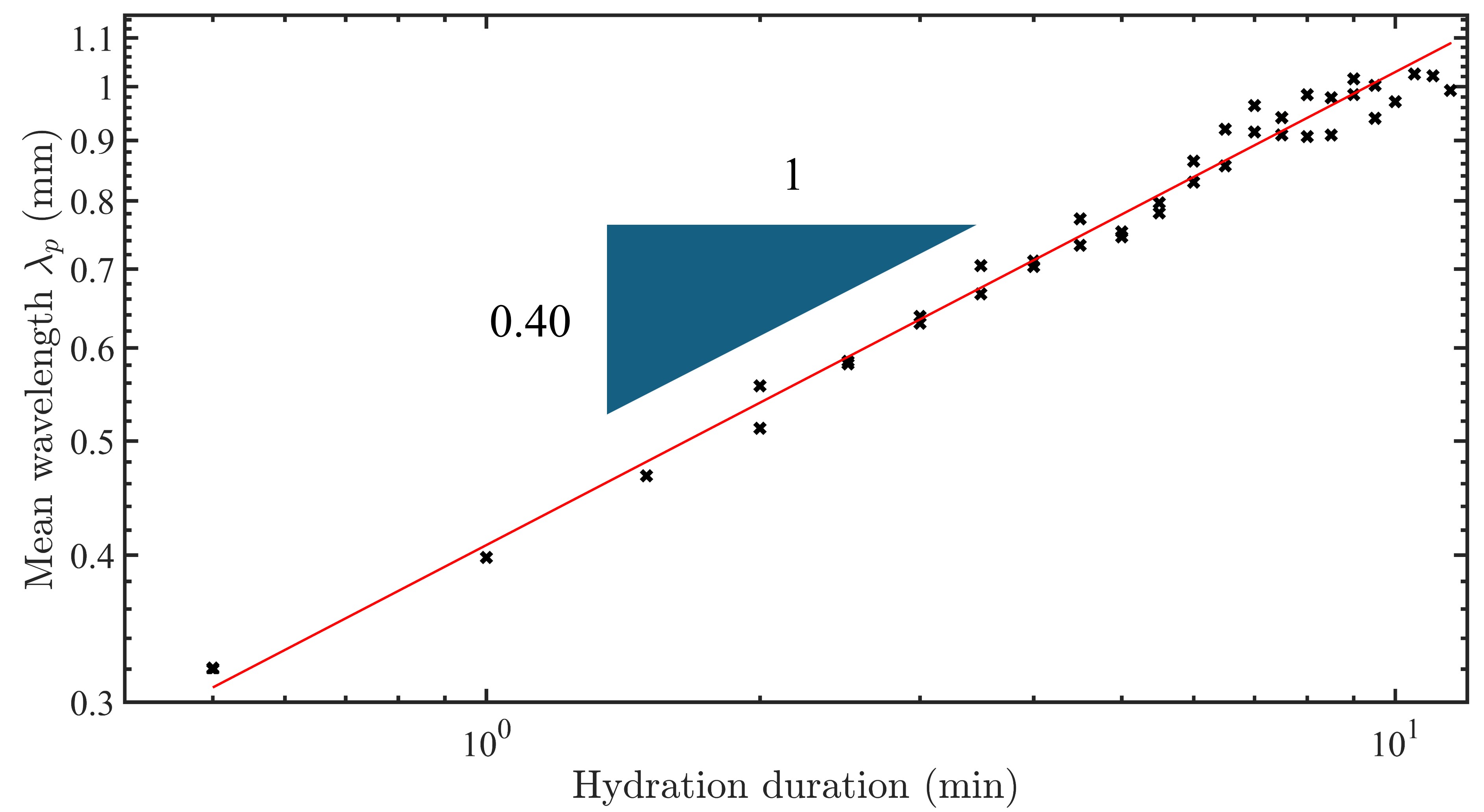}
	\caption{Log--log plot showing the mean wavelength of the pattern  as a function of hydration duration for experiments with a fixed water drop volume of \SI{100}{\mu\litre}. The red line is the best least-squares  fit.
	}
	\label{fig:Fortune8}
\end{figure}

We study the impact of the hydration duration, namely the time between when the water is added to the system and when the acetone is added to the system, on the pattern of the instability. Fig. \ref{fig:Fortune6} presents photographs of the pattern for water drops of volume \SI{100}{\micro\litre} for a range of different hydration durations from 1 min to 20 min. All images were taken with the acetone layer still above the hydrogel sheet. The \SI{100}{\micro\litre} drop volume was chosen as a representative drop size since it results in blisters that are an order of magnitude larger than the wavelength of the cells formed, whilst the edges of the drop are far away from the edges of the hydrogel sheet. Based on the results presented in \S\ref{ConstantdehydrationTime} the structure of the pattern is independent of blister size. Hence, if we used a different droplet volume than \SI{100}{\micro\litre}, we would still make the same observations as using the images in Fig. \ref{fig:Fortune6}.

For hydration durations of less than 15 minutes (Figs. \ref{fig:Fortune6}(a)--(g)) the pattern shows a clear regular structure. The wavelength of this instability increases as a monotonic function of the hydration duration. The cells on the outside of the blister are typically slightly smaller than the interior cells and form a ring (denoted in Fig. \ref{fig:Fortune6} by blue dots). These patterns are qualitatively similar to those observed in Appendix B of \citet{Etzold23} using a dyed (\SI{0.01}{\percent} by weight methylene blue) droplet. However, a quantitative comparison between the methylene blue patterns and our patterns is not possible. Methylene blue is a cationic dye and thus we believe will induce structural changes in the hydrogel arising from ionic interactions between polar functional groups \cite{Nadtoka20}.  

Fig. \ref{fig:Fortune7} presents a box plot of the distribution of the pattern wavelength for a set of sixteen experiments with a constant initial water volume of \SI{100}{\micro\litre} but different hydration durations using hydrogel coupons that came from the same initial hydrogel sheet. A statistically meaningful increasing trend can be seen. In Fig. \ref{fig:Fortune8}, utilising a larger dataset of 36 experiments, we find that the mean pattern wavelength $\lambda_p$ can be fitted with a power law function of the hydration duration, as $\lambda_p \sim t^{\alpha}$ where $\alpha \approx 0.40$ whose \SI{95}{\percent} confidence interval is $[0.38, \, 0.42]$. 

At a hydration duration of approximately 15 minutes (Fig. \ref{fig:Fortune6}(h), the pattern structures becomes less regular as the cells become longer and thinner, and starts to disappear. After 20 minutes, the pattern is only just visible to the naked eye, with only a few small optically opaque regions visible, which are morphologically different to the cells observed for shorter hydration durations.  

\subsection{Swelling in Water} \label{WaterSwelling}

\begin{figure}[!tb]
	\centering\includegraphics[trim={0 0cm 0 0cm}, clip, width=\columnwidth]{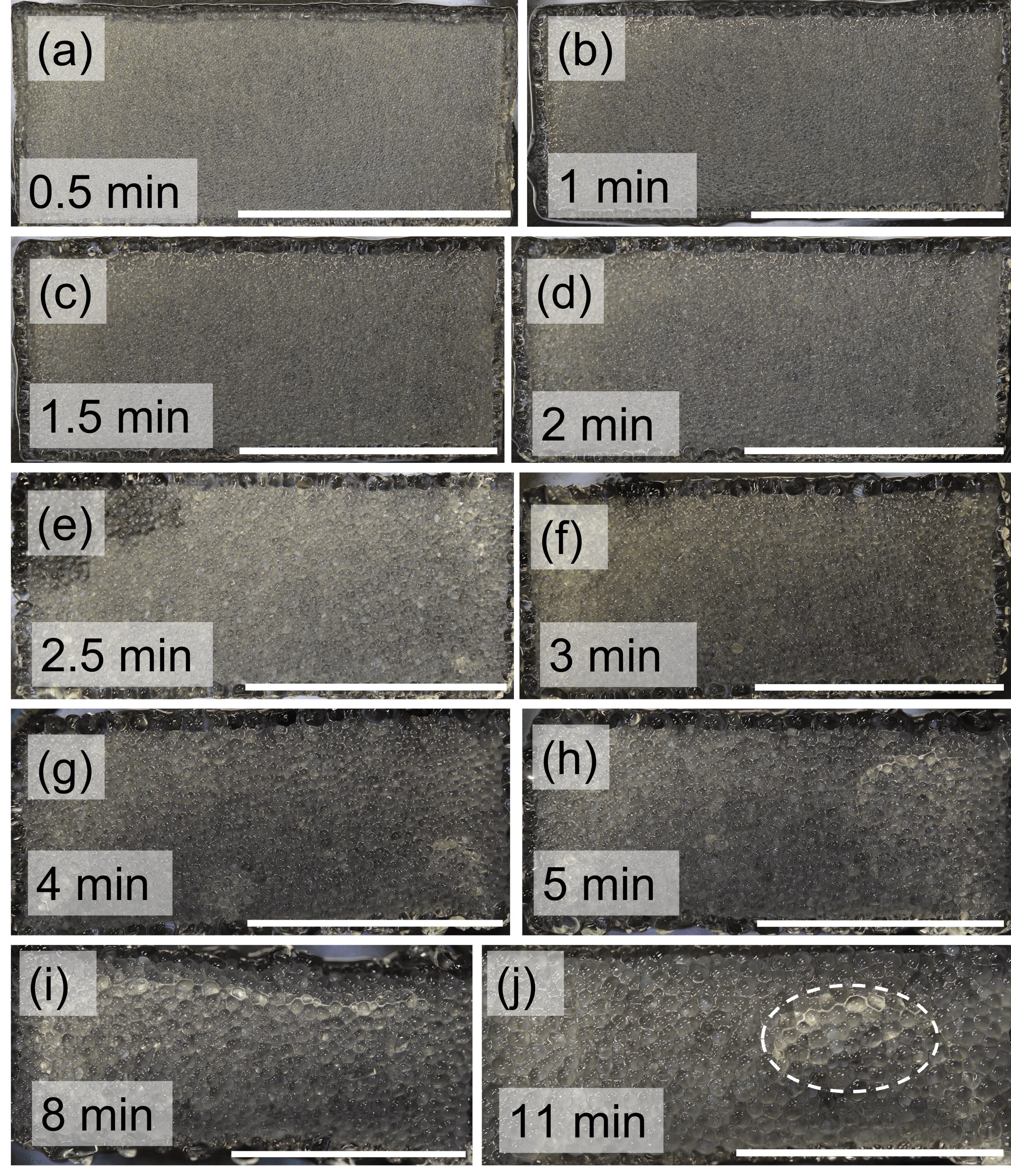}
	\caption{Sequence of photographs showing the patterns that emerge when a hydrogel sheet is swollen in a water bath and imaged every thirty seconds following the protocol of $\S \ref{waterdunking}$. (a--j) The wavelength of the buckling instability increases with time. (g--j) At later times, the whole hydrogel sheet buckles out of the plane. A white dashed ellipse encircles a region that has buckled out of the plane in (j). White scale bars indicate 20mm.
	}
	\label{fig:Fortune9}
\end{figure}
\begin{figure}[!tb]
	\centering\includegraphics[trim={0 0cm 0 0cm}, clip, width=\columnwidth]{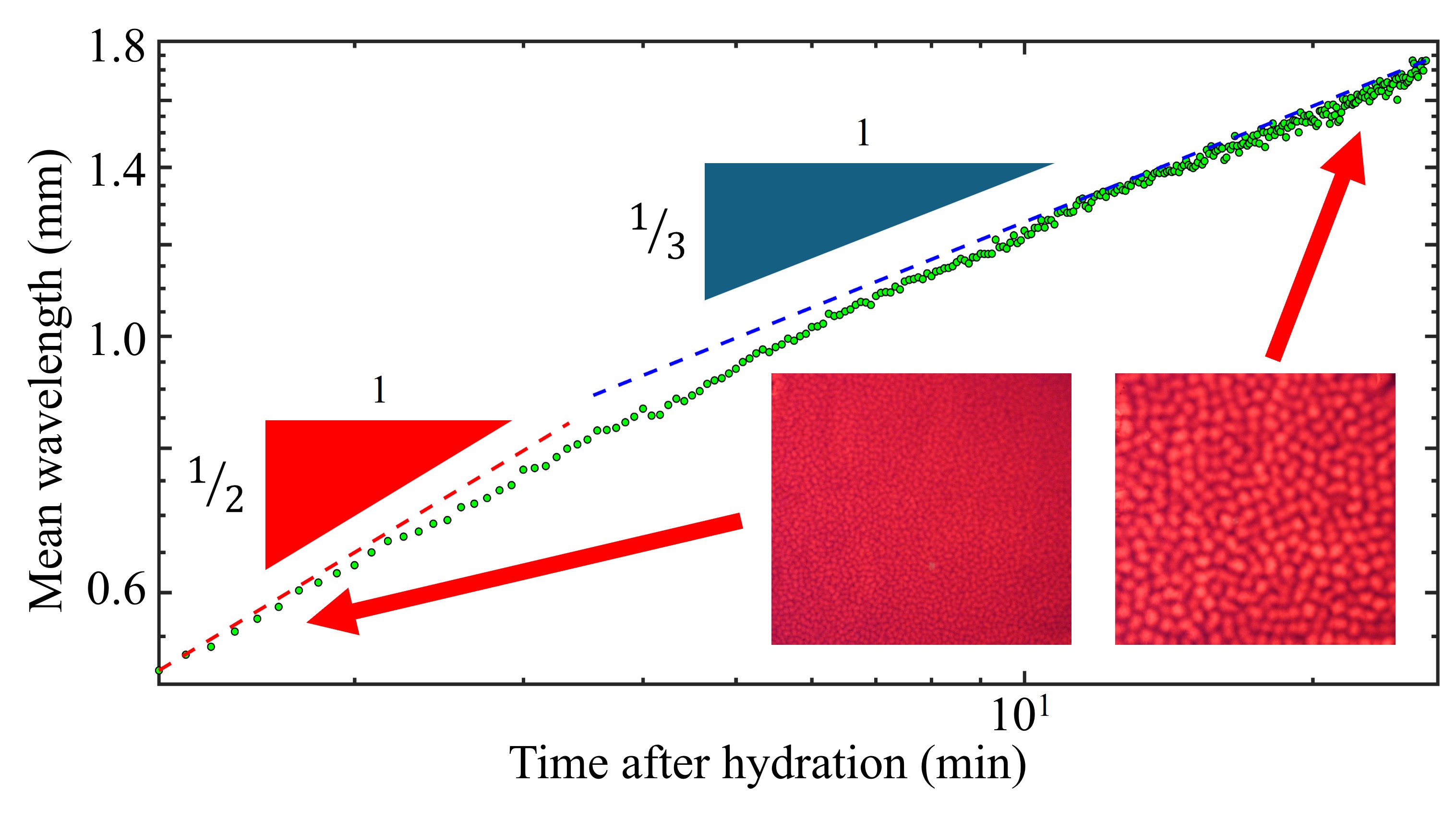}
	\caption{The mean wavelength of the crumpling instability in water swollen hydrogel plotted as a function of time after hydration. The two insets show representative experimental images at early and late times, respectively, taken using the experimental protocol given in $\S \ref{waterdunking2}$. Vertical error bars are estimated to be $\approx \pm 0.1$ mm.
	}
	\label{fig:Fortune10}
\end{figure}

To explore why the wavelength $\lambda_p$ of the pattern grows temporally like $t^{0.40}$, we return to the original crumpling instability when the hydrogel sheet is swelling in water. Fig. \ref{fig:Fortune9} presents the patterns that emerge when a hydrogel sheet is swollen in a water bath and imaged by removal every thirty seconds, following the protocol detailed in \S\ref{waterdunking}. The variable lighting inherent in some of the images (for example the top left corner of Fig.\ref{fig:Fortune9}(e)) arose from reflections from residual water below the hydrogel sheet. The edges of the sheets swelled considerably more than the interior since the swelling of the polymer at the droplet edge is less constrained than in the centre. This edge swelling led to some detachment of gel and thus mass loss each time that the hydrogel sheets were removed from the water bath for filming. 
Furthermore, due to elastic stresses created in the swelling process, the hydrogel sheets buckle out of the plane. This is particularly apparent at late times in Figs. \ref{fig:Fortune9}(g--j) (a region of uplift is encircled with a white dashed ellipse in (j)).

Qualitatively, these patterns are similar to those observed above in the acetone experiments. This provides some evidence that the  patterns observed after dehydration with acetone are not affected by the addition of the acetone, but are simply `locked-in-place'. An interesting difference, however, is that the bumps formed by the crumpling instability in water swollen hydrogel have rounded edges. In contrast, the cells formed after dehydration with acetone are polygonal with sharp angular corners. A possible explanation is that the hydrogel has undergone a phase change from a rubbery state to a more glassy state with corresponding changes in material properties, and thus the gel has become much stiffer \cite{Zhang19}. This stiffness allows the gel to maintain angular corners. This transition from round bumps to angular corners can be seen in many other elastic systems experiencing deformation. For example, \citet{Chiang21} explores the crumpling of thin elastoplastic shells. As the applied pressure increases, the crumpling pattern evolves from rounded craters to polygonal ones with this transition occurring at lower pressures for thicker (and hence stiffer) shells.

Quantitatively, Fig. \ref{fig:Fortune10} plots the mean wavelength of the crumpling instability in water swollen hydrogel as a function of the time after hydration, following the protocol detailed in \S\ref{waterdunking2}. As can be seen, at early times the crumpling wavelength grows with power law behaviour as $t^{1/2}$. The growth rate decreases as a function of time, tending towards a $t^{1/3}$ dependence at later times. Hence, taking this information and comparing it with the acetone dehydration experiments, we see that in the dehydration experiments the acetone is added at intermediate times when the crumpling instability in the water swollen hydrogel has a wavelength that is growing like $t^{\alpha}$ with $\alpha \in [1/3, \, 1/2]$. Hence, since the growth rate measured for the experiments with acetone immersion is $0.40 \in [1/3, \, 1/2]$, this is further evidence that the wavelength of the acetone induced pattern is set by the wavelength of the original crumpling instability. 
\section{Discussion and Conclusions}\label{conclusions}

We have presented a novel pattern formation phenomenon that we have observed after dehydrating a hydrogel sheet swollen by water through static immersion in acetone. We now discuss a hypothesis for the underlying physical mechanism behind this acetone induced pattern formation through comparison with experiments capturing the original crumpling instability in water swollen hydrogel.

From \citet{Zhang19}, it is known that when a hydrogel is immersed completely in a poor solvent (e.g. acetone), it shrinks and becomes turbid due to phase separation. Due to cross-links between polymer chains, this phase separation is highly localised, occurring on the microscale rather than on the macroscale. More precisely, as water is removed from the hydrogel, interactions between polymer chains gradually collapse the polymer network. This collapse leads to the inner gel morphology transitioning from a porous-medium-like framework to a bi-continuous domain structure consisting of regions with a dense polymer concentration and regions with a sparse polymer concentration. The turbidity is caused by the dense polymer domains (denoted from now on as inhomogeneities) whose size is comparable to the wavelengths of visible light \cite{Khandai20}.

\begin{figure}
	\centering\includegraphics[trim={0 0cm 0 0cm}, clip, width=\columnwidth]{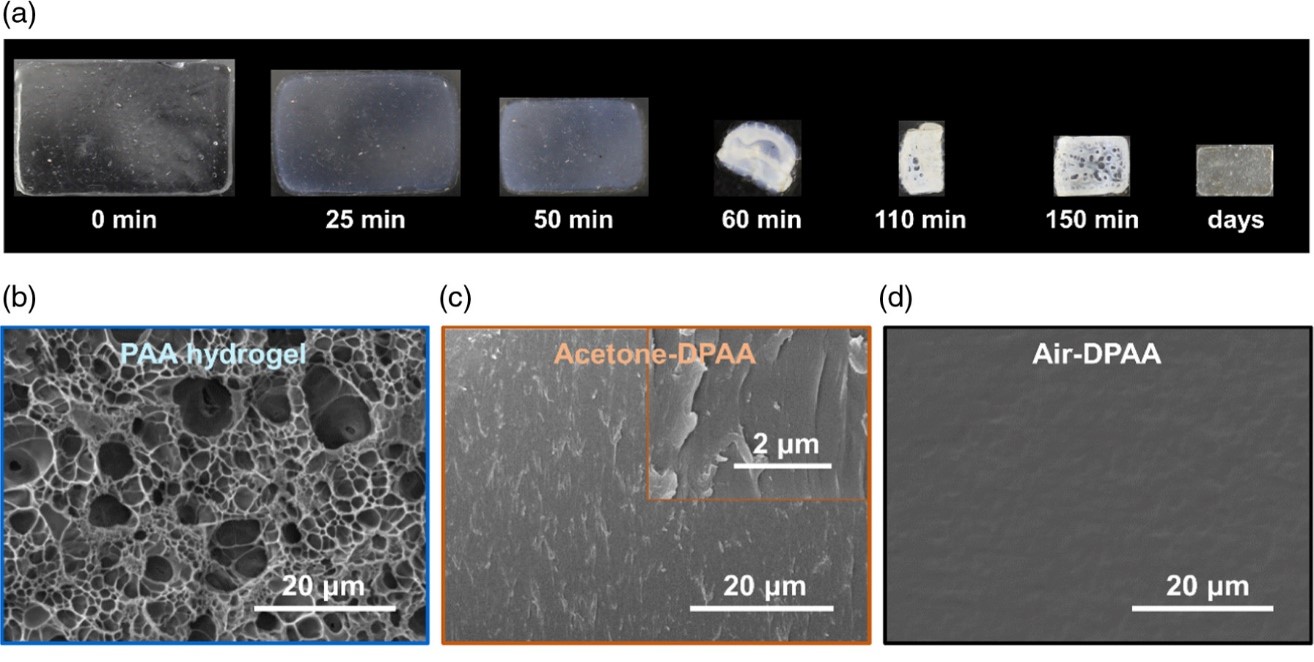}
	\caption{Dehydration of a hydrogel cuboid through complete immersion in acetone. (a) The hydrogel shrinks as it undergoes phase separation, turning from transparent to opaque and then back to transparent. (b) The cross-sectional morphology of freeze-dried PAA hydrogel showing porous structures (corresponding to hydrated hydrogel). (c) The cross-sectional morphology of acetone dehydrated PAA hydrogel showing flaky lamella structures (corresponding to acetone dehydrated hydrogel). (d) The cross-sectional morphology of air dried PAA hydrogel showing a lack of microstructures (corresponding to air dehydrated hydrogel). Reproduced and adapted from figure 1 of \citet{Zhang19}.
 }
	\label{fig:Fortune11}
\end{figure}

As an illustration of this phase transition behaviour from a transparent rubbery state to a turbid glass state, Fig. \ref{fig:Fortune11} reproduces panels (a--d) of figure 1 of \citet{Zhang19}. A hydrogel cuboid is fully immersed in acetone. Losing water into the acetone, it shrinks and undergoes a phase transition over time, transitioning from transparent to turbid and then back to transparent (see Fig. \ref{fig:Fortune11}a). Initially, the hydrogel has a homogeneous pore structure (see Fig. \ref{fig:Fortune11}b) and thus is optically transparent. Then, as the hydrogel transitions to a bi-continuous domain structure, inhomogeneities form on a length scale comparable to visible light, leading to refractive index changes. These refractive index changes lead to the scattering of visible light and thus turbidity (see Fig. \ref{fig:Fortune11}(c)). Finally, after being left for days in acetone, the bi-continuous domain structure becomes more homogeneous, leading to less visible light scattering and thus a return to transparency (see Fig. \ref{fig:Fortune11}(d)). 

These observations provide further insight about the acetone-induced pattern formation. Considering a typical acetone dehydration experiment, at the point at which the acetone is added, the hydrogel surface is exhibiting a transient crumpling instability comprising of a series of bumps that intersect at depressions or creases. When the swollen hydrogel is immersed in acetone, differences in chemical potentials produce large pressure gradients that drive water out of the hydrogel along with a small ingress of acetone.  This sudden egress of water can collapse the polymer network faster than the stretched polymer chains can relax, triggering a phase transition with the formation of inhomogeneities that turn the cell optically turbid. 

From past work in the literature (e.g. \cite{Tanaka87}) it is  known that the polymer fraction in the creases is considerably larger than in the cells. 
After dehydration, less water egresses and less acetone ingresses the creases. Less polymer collapse occurs leading to fewer inhomogeneities forming and thus the creases remaining transparent. Furthermore, the outer regions of a cell are more turbid than the cell interior because they correspond to a larger surface area when the cell was a fully three-dimensional structure before dehydration and thus a larger water egress. Away from the blister, only a small amount of water is removed from the hydrogel and thus polymer collapse is not initiated. The hydrogel does not undergo a phase transition but instead remains transparent. 

The acetone induced cells appear more angular than the water-induced crumpling instability bumps. In the water swelled hydrogel, the polymer chains are highly solvated, screening polymer-polymer interaction.  As a result, the hydrogel behaves elastically, leading to approximately circular bumps. In contrast, after phase separation, the polymer chains are close enough together to form a locally dense phase with strong polymer--polymer interactions \cite{Sato15}. Thus, the induced extra tensile strength is a possible cause of the observed angularity at the corners of the cells.

This pattern formation phenomenon highlights the fact that while it is convenient to think about swelling polymers as deformable poroelastic networks, they can display much more complicated behaviours due to the presence of multiple phases. For example when decontaminating a medium by chemically removing bulk fluid, we cannot simply assume that the morphology of the polymer remains constant throughout the process. Small initial perturbations can trigger phases transitions leading to a  bi-continuous domain structure with very different optical, mechanical and chemical properties. 

More generally, we note that by controlling the time before dehydration we can control reliably the wavelength of the instability. Furthermore, we have shown that the morphology is independent of the total size of the blister containing the pattern. Hence, by adding a range of water drops at different times to the hydrogel, we could potentially pattern the sheet with arbitrary design of regions with cells of  different sizes. For example, placing a large drop in the centre of the hydrogel sheet, waiting several minutes, then adding a ring of smaller drops around the first drop before waiting a further one minute until dehydration would recover a central region of large cells surrounded by small cells. From \citet{Zhang19}, we note that the glassy phase transition dramatically changes the mechanical properties of the hydrogel to a material that is flexible with a higher tensile strength and a more pronounced self-healing capabilities. Hence, this pattern formation could be exploited to manufacture a material with spatially varying mechanical properties. 

\section*{Author Contributions}
MAE, SBD and JRL identified the problem's practical application and acquired the funding. GTF identified the pattern formation, developed the experiments and performed the subsequent image analysis with input from MAE, SBD and JRL. GTF prepared the first draft of the manuscript. All authors contributed on further revisions of the manuscript.

\section*{Conflicts of interest}
There are no conflicts to declare.

\section*{Acknowledgements}
GTF, MAE, JRL and SBD acknowledge funding from the Defence Science and Technology Laboratory (Dstl).

\bibliography{Fortune_et_al}
\bibliographystyle{rsc}

\end{document}